%% file: author.tex
\documentclass{svproc}
\usepackage{cite}
\usepackage{acronym}
\usepackage{xcolor}
\usepackage{fancybox}
\usepackage{enumitem}
\usepackage{tikz}
\newcommand{\circlednumber}[1]{%
    \tikz[baseline=(char.base)]{
        \node[shape=circle, draw=black, fill=black, text=white, inner sep=0.7pt, font=\sffamily\footnotesize] (char) {#1};
    }
}

\usepackage{url}

\usepackage{longtable}

\newcommand\blfootnote[1]{%
  \begingroup
  \renewcommand\thefootnote{}%
  \footnote{\footnotesize #1}
  \addtocounter{footnote}{-1}%
  \endgroup
}

\begin{document}

\input{acronym}

\mainmatter              
\title{Programming in Brazilian Higher Education and High School: A Systematic Literature Review}
\titlerunning{Programming in Brazilian Education: A Systematic Literature Review}  
%
\author{Sofia C. {Latini Gonçalves}\inst{1} \and Rodrigo Moreira\inst{1} \and Larissa F. {Rodrigues Moreira}\inst{1} \and Andr\'{e} R. Backes\inst{2} \and Adriana Zanella Martinhago\inst{1}}
\authorrunning{Sofia C. Latini Gonçalves et al.} 
%
%
\institute{Institute of Exacts and Technological Sciences\\ Federal University of Viçosa (UFV), Rio Paranaíba/MG, Brazil\\
\email{sofia.goncalves, rodrigo, larissa.f.rodrigues, adriana.martinhago@ufv.br}\\ 
\and
Department of Computing (DC)\\ Federal University of São Carlos (UFSCar), São Carlos/SP, Brazil\\
\email{arbackes@yahoo.com.br}
}

\maketitle              

\begin{abstract}
\blfootnote{\color{red}This paper was accepted and published in the proceedings of the 19th Latin American Conference on Learning Technologies (LACLO 2024).}
Programming, which is both economically significant and mentally stimulating, has been found to benefit the aging brain and to enhance cognitive function at various educational levels. Despite its advantages, challenges persist in standardizing and implementing programming education effectively across both the higher and secondary education levels in Brazil. To shed light on these issues, we carried out a systematic review of programming teaching methods in the Brazilian context, examining gaps, common techniques, approaches, and action opportunities in programming education. Our findings provide valuable recommendations for educational policymakers and educators to develop effective and updated national policies to teach programming.
\end{abstract}

\section{Introduction}\label{sec:introduction}

Digital transformation positively impacts various areas of our society, enhancing how we conduct business, learn, and communicate~\cite{Mukul2023}. Behind this behavioral shift lies disruptive computational technologies, such as integrated nano-circuit programming, Artificial Intelligence, Computer Vision, and pervasive communication~\cite{Pivoto2023, Larissa2023}. It is important to note that programming is a key element of digital transformation progress, and its use is necessary for the implementation of solutions that are currently in use and for the development of new applications that are becoming increasingly robust and user-oriented.

Programming is a curricular component present at different stages of a computing professional's education. There are also initiatives that have incorporated programming at even higher elementary levels of education, such as high school or elementary school. During these phases, the curriculum often emphasizes problem solving and computational thinking alongside the standard student journey curriculum. In developed countries, the incorporation of computer programming or coding is part of the mathematics curriculum, highlighting the benefits of such content at these stages of education~\cite{Holo2023}. However, in developing countries, such as Brazil, there is still a lack of standardization in the approaches to computer programming content, especially in elementary education, which presents challenges to be overcome~\cite{Storte2019}.

Therefore, we investigated how different techniques are used across different institutions and geographic regions. We conducted a systematic review analyzing studies to understand and address research questions related to programming teaching within the Brazilian context. Our literature analysis innovatively sheds light on specific insights into programming teaching, such as the primary coding languages used, the methods employed, and the target domain. Understanding these aspects enables decision makers to develop robust and current national education policies. 

Among the contributions of this paper, we highlight the identification of gaps in programming education within the Brazilian context, an investigation into the commonly used methods across different educational levels, and an analysis of the diverse approaches adopted nationwide to address the challenges of programming teaching.

The remainder of this paper is organized as follows. Section~\ref{sec:related_work} contrasts previous attempts to systematize research results on the theme of programming teaching. Section~\ref{sec:evaluation_method} details the approach used in the literature survey. Section~\ref{sec:evaluation_method} presents the systematic method adopted, and Section~\ref{sec:results_and_discussion} presents the answers and insights from our analyses. Finally, in Section~\ref{sec:concluding_remark} we remark on some concluding points and call for action in future work.

\section{Related Work}\label{sec:related_work}

In this section, we provide a short state-of-the-art description of previous studies that aimed to explore how programming education is used in high school and higher education, along with its implications. We analyzed related works, highlighting their study objectives and contrasting them with the proposed study. Each work examined specific challenges faced by students and educators, as well as the strategies and methodologies employed to enhance programming teaching and learning.

Duarte de Holanda et al.~\cite{DuartedeHolanda2019} conducted a study to understand the current research landscape of scientific publications in Brazil regarding strategies employed in Programming education in higher education from 2014 to 2018. The findings indicate the development and utilization of educational software, digital games, and specific methodologies aimed at mitigating the failure rates in these disciplines. 

Medeiros et al.~\cite{Medeiros2019} investigated the challenges in introductory programming courses, and their review included studies up to 2016. They found that students faced issues related to problem-solving abilities, mathematical knowledge, motivation, engagement, and syntax in programming languages. The faculty members highlighted the need for appropriate teaching methods and tools. Subsequently, ~\cite{Medeiros2020} discussed the learning and teaching of Programming in Brazilian higher education institutions from 2010 to 2016. Despite advancements in methods and tools, high dropout and retention rates have persisted. Mehmood et al.~\cite{Mehmood2020} also analyzed the challenges in learning programming and analyzed studies published between 2011 and 2020. They concluded that the pedagogical approach, programming language choice, and student performance analysis were commonly addressed, whereas curriculum content, assessment tools, and tool-based learning were less explored. 

Morais et al.~\cite{Morais2020} conducted a study  to identify the main difficulties and challenges encountered by higher education students in introductory programming courses and their impact on academic performance. The main difficulties encountered were related to a lack of prior knowledge in mathematics, dissociation between the proposed exercises and real-world applications, and challenges in understanding the subject matter, often resulting in course failure, dropout, and frustration. Reis et al.~\cite{Reis2020} carried out a systematic review to identify how robotics can be applied to teaching programming and investigated how published studies used electronic devices based on free hardware. In addition, the authors propose a flowchart for classifying robotic devices and electronic circuits that can be used in programming teaching. Reis da Silva et al.~\cite{ReisdaSilva2021} identified that between 2014 and 2020, the majority of studies carried out in Brazil were dedicated to teaching programming in elementary education using the Scratch tool, highlighting the need for research and advances in high school and higher education.  

Calderon et al.~\cite{Calderon2021} identified the main active methodologies employed by Brazilian higher education educators between 2010 and 2021, with educational games and gamification being widely used to promote increased student participation and engagement in activities. Moraes et al.~\cite{Moraes2022} analyzed studies published between 2016 and 2020, encompassing Brazilian conferences and journals. They identified numerous proposals for hybrid or online programming teaching to enhance student motivation alongside the development of tools for automated code correction to assist instructors. Castro and Santos~\cite{CastroSantos2023} studied the impact of gamification in higher education courses in the subjects of programming logic, and gamification proved to be a promising approach and capable of making the learning process more engaging.

This study differs from previous studies in that it selects and analyzes papers that present methodologies, various programming languages, and techniques related to teaching programming in Brazilian high school and higher education published between 2013 and 2023. In addition, our study identified the best strategies to help overcome dropout problems in higher education.

\section{Systematic Review Method}\label{sec:evaluation_method}

The method employed in this study is based on a systematic review, which constitutes a type of secondary analysis used to gather published works in the literature (primary studies), aiming to obtain a broad set of information on a central research topic~\cite{Petersen2008}. The process of conducting this review was divided into five steps: (i) defining the research questions, (ii) defining the search string, (iii) establishing inclusion and exclusion criteria, (iv) selecting relevant primary studies, and (v) data extraction and analysis of studies.

\subsection{Research Questions}\label{sec:RQ}

We defined the following Research Questions ($\mathcal{RQ}s$):

\begin{itemize}[label=$\bullet$]
    \item $\mathcal{RQ}_{1}$: What are the overview of studies in Brazil?
    \item $\mathcal{RQ}_{2}$: Which methods have been employed in programming teaching proposals?
    \item $\mathcal{RQ}_{3}$: Which programming content should be addressed the most?
    \item $\mathcal{RQ}_{4}$: How are programming teaching methods evaluated?
    \item $\mathcal{RQ}_{5}$: How does the programming language integrate into the programming teaching method?
    
\end{itemize}

\subsection{Search Procedure}

After defining $\mathcal{RQ}s$, the next step is to construct a search string. We conducted a search for candidate papers using the SBC-OpenLib (SOL)\footnote{Available in: \url{https://sol.sbc.org.br}}, the Digital Library of the Brazilian Computing Society, considering publications in proceedings from 2013 to 2023 to cover the last 10 years, and employing specific keywords as the search string. We chose to exploit the SOL database because it contains important published works and is representative of the Brazilian context. We applied the following search string. \bigskip

\setlength{\fboxsep}{7pt}
\shadowbox{\parbox{0.88\linewidth}{
\footnotesize{\ttfamily{("Programming" OR "Programação") AND ("Ensino" OR "Learning" OR "Algorithms" OR "Higher Education" OR "Coding Class")}}
}}

\subsection{Defining Inclusion and Exclusion Criteria}

Papers included in the review had to be written in English or Portuguese and published in proceedings available in SOL database on the theme of programming teaching in higher education and high school. We defined the following criteria for inclusion and exclusion in the selection of relevant studies.

\begin{itemize}
    \item \textbf{Inclusion Criteria:}

    \begin{itemize}[label=$\circ$]
        \item Papers on methods of teaching programming in higher education and high school contexts.
    \end{itemize}

    \item \textbf{Exclusion Criteria:}

    \begin{itemize}[label=$\circ$]
        \item Duplicated papers;
        \item Papers whose language was not English or Portuguese;
        \item Papers not from Brazil;
        \item Secondary studies published in the literature;	
        \item Papers that do not specify the teaching method;
        \item Papers that have not been applied in high school or higher education;	
        \item Papers that do not focus on teaching and/or learning programming;	
        \item Papers proposing the presentation of programming teaching environments.
    \end{itemize}
\end{itemize}

\subsection{Selecting Relevant Primary Studies}

\begin{figure}[!htbp]
  \centering
  \includegraphics[width=\linewidth]{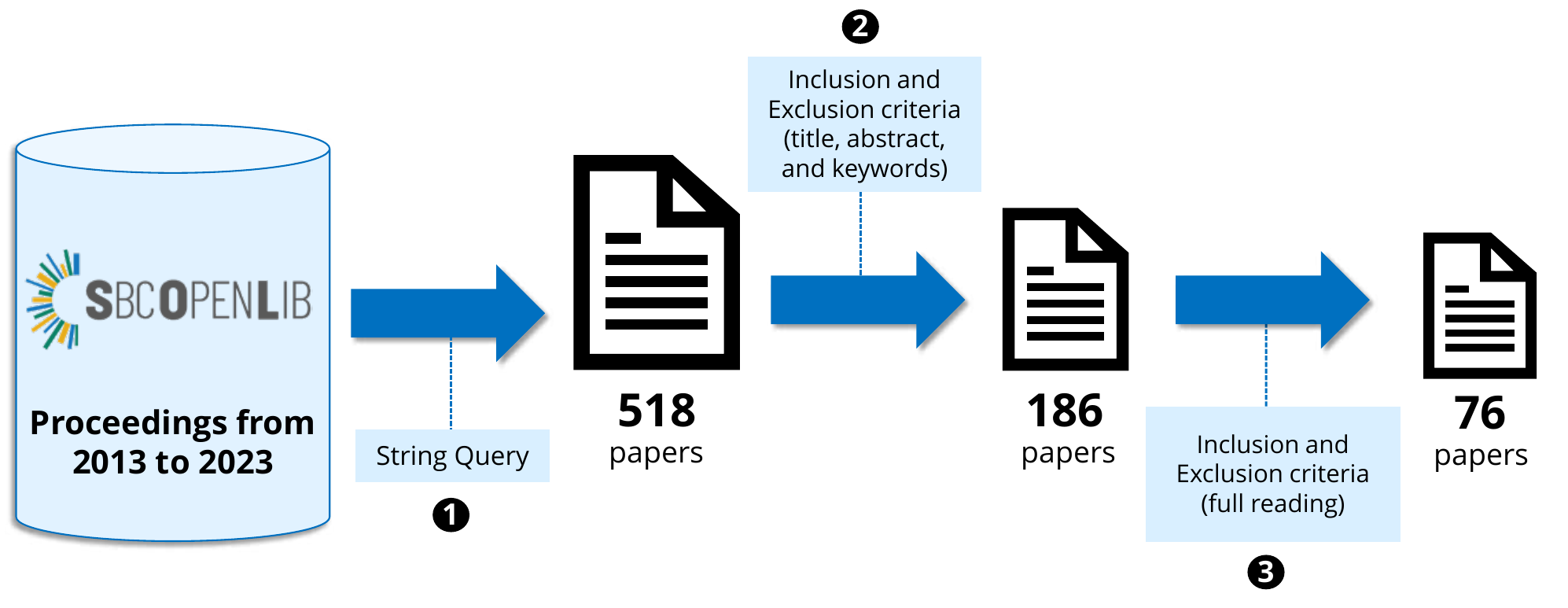}
  \caption{Paper selection process.}
  \label{fig:method}
\end{figure}

Figure~\ref{fig:method} illustrates the paper selection process, which is divided into three (3) steps, as described as follows. 

\circlednumber{1} \textbf{First Step:} we applied the search string and obtained 518 primary studies from SOL database. In addition, we evaluated papers published in English and Portuguese. The papers obtained in the first step were retrieved for the analysis in the second step.

\circlednumber{2} \textbf{Second Step:} the 518 papers were randomly divided among the researchers (authors of this work) and subjected to inclusion and exclusion criteria based on the assessment of the title, abstract, and keywords. Subsequently, the number of papers was reduced to 186. 

\circlednumber{3} \textbf{Third Step:} the 186 papers were again randomly divided by the researchers for a complete reading and once again the inclusion/exclusion criteria were applied. After this process, 76 relevant papers were selected for data extraction.

\subsection{Protocol Validation}

This review outlines the typical challenges encountered in systematic reviews, particularly regarding search coverage and potential biases during the study selection, data extraction, and analysis stages. One limitation of this study is the need for answers to $\mathcal{RQ}s$ that may not have clear-cut or objective responses. To mitigate these challenges, we adhered to the best practices for systematic reviews and involved multiple researchers at various stages, such as study selection, quality evaluation, and data extraction~\cite{Keele2007, Petticrew2008}.

\section{Results and Discussion}\label{sec:results_and_discussion}

This section presents the answers to the $\mathcal{RQ}s$ raised in Section~\ref{sec:RQ} based on 76 relevant studies\footnote{Available in: \url{https://t.ly/ZpUYr}}.

\textbf{$\mathcal{RQ}_{1}$: \textbf{What are the overview of studies in Brazil?}}

\begin{figure}[!htbp]
  \centering
  \includegraphics[width=0.6\linewidth]{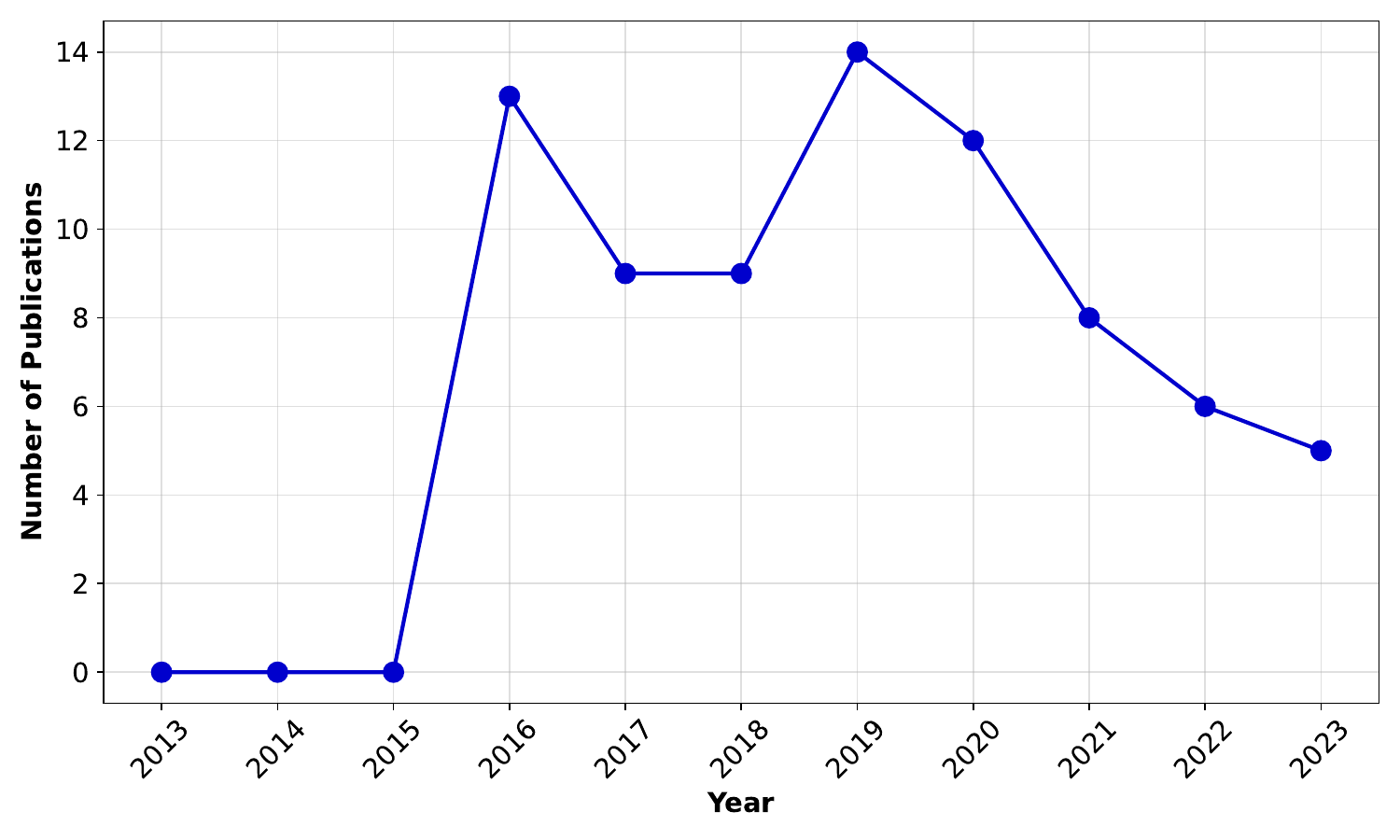}
  \caption{Number of publications per year.}
  \label{fig:publications-year}
\end{figure}

To answer $\mathcal{RQ}1$, initially, our aim was to comprehend the publication trends regarding computer programming teaching strategies at intermediate and higher education levels. Figure~\ref{fig:publications-year} provides an overview of the distribution of the 76 papers arranged by the year of publication from 2013 to 2023. 

However, since 2021, we have observed a significant reduction in the number of studies published on programming teaching in higher and secondary Brazilian education. This decrease may have important implications for the dropout problem in Science, Technology, Engineering, and Math (STEM). While the reduction may reflect changes in research priorities or external factors, such as the global pandemic, it also raises questions about the quality and effectiveness of programming teaching methods. In particular, high dropout rates in Computer Science courses are a growing concern, and the lack of innovation and adaptation in teaching approaches can contribute to student demotivation and a loss of interest in the field. Therefore, a decrease in the volume of publications may indicate a gap in research and intervention to develop more effective and engaging teaching strategies.

\begin{figure}[!htbp]
  \centering
    \centering
    \includegraphics[width=0.7\linewidth]{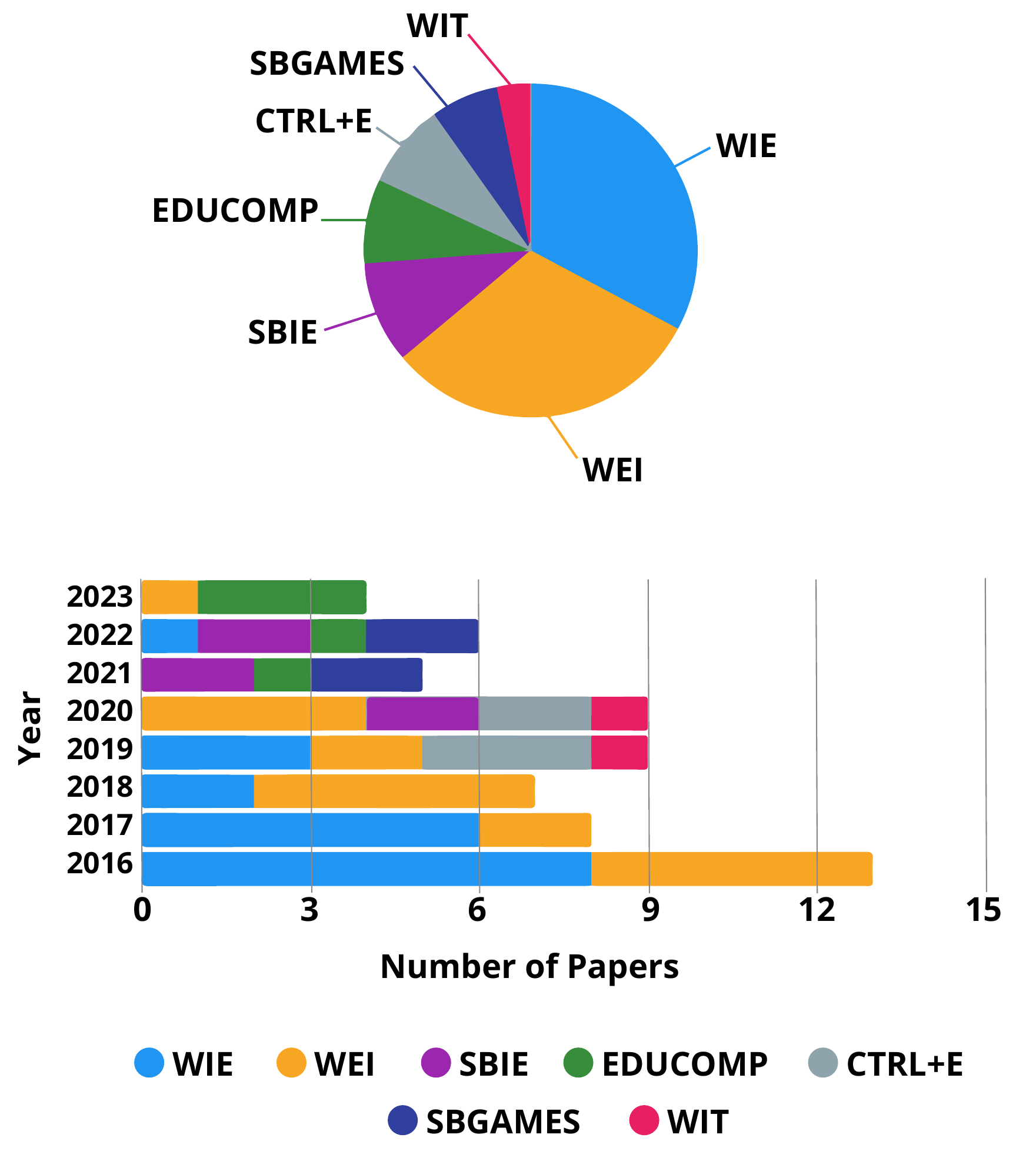}
    \caption{Allocation of publications over years in contrast to conferences.}
    \label{fig:chart1}
\end{figure}

We also analyzed how publications were distributed over the years at the conferences. As shown in Figure~\ref{fig:chart1}, the conferences that received the most papers on this topic were Workshop de Informática na Escola (WIE), Workshop on Computing Education (WEI), Brazilian Symposium on Computers in Education (SBIE), Simpósio Brasileiro de Educação em Computação (EDUCOMP), Congresso sobre Tecnologias na Educação (Ctrl+E), Brazilian Symposium on Computer Games and Digital Entertainment (SBGames), and Women in Information Technology (WIT).

From 2016 to 2018, publications predominantly occurred at WIE and WEI conferences. However, starting in 2019, other conferences began to receive papers on the topic, such as WIT, SBIE, and Ctrl+E. It is noteworthy that among the selected studies, no papers were obtained from WIE or WEI in 2021, suggesting that potential publications from this event may have migrated to other conferences held in the same year (EDUCOMP, SBIE, and SBGAMES). Furthermore, from this analysis, we noticed that from 2021 onwards, the first efforts with themes associated with digital games for programming education in secondary and higher education emerged, as evidenced by publications in SBGAMES, as exemplified in~\cite{Oliveira2021, Silva2022}.

\begin{figure}[!htbp]
    \centering
    \includegraphics[width=0.55\linewidth]{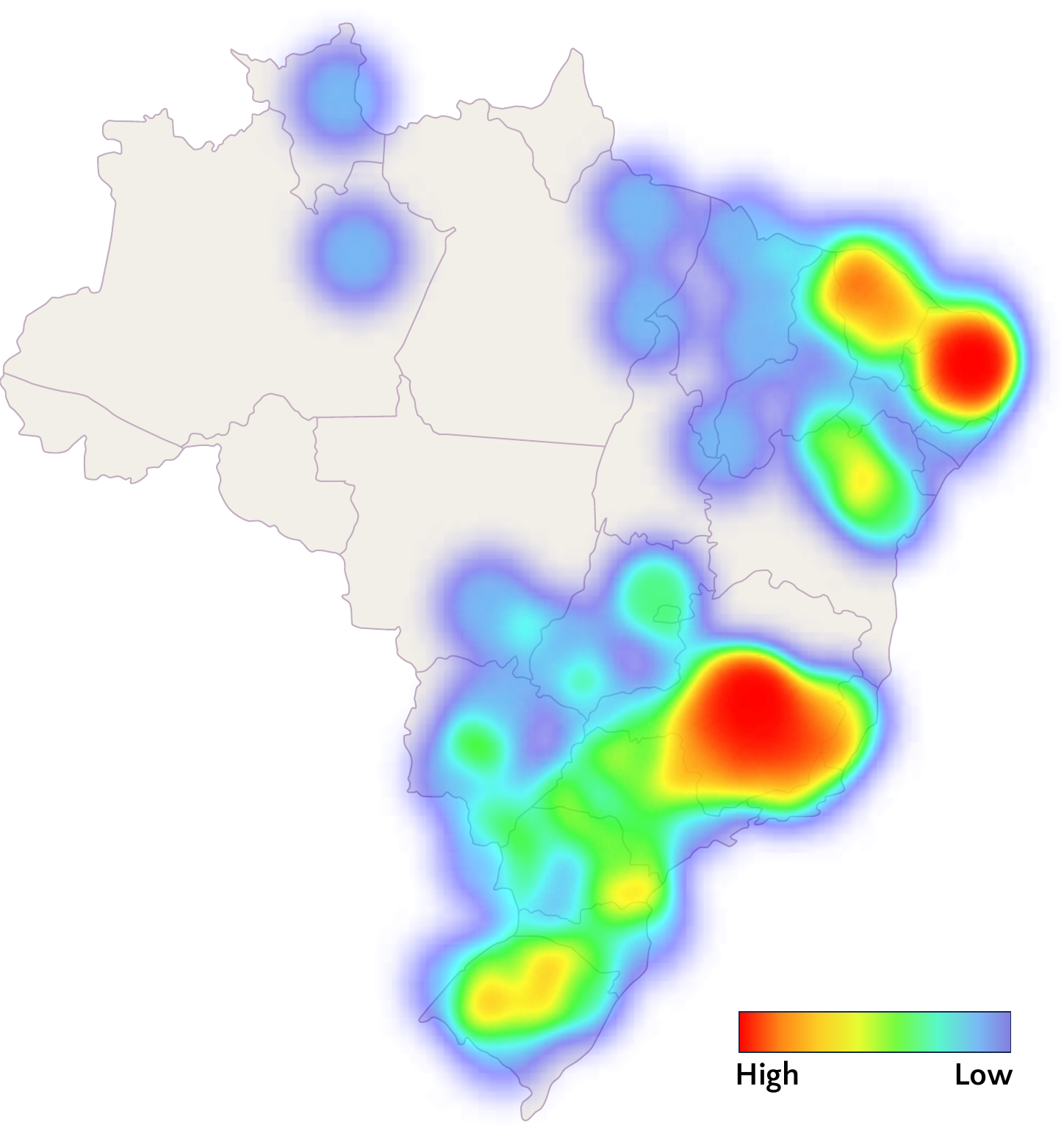}
    \caption{Heat map of the distribution of publications across regions of Brazil.}
    \label{fig:map}
\end{figure}

Table~\ref{tab:publication_distribution} presents the distribution of publications across different regions of Brazil, along with the research conducted in international collaboration with Portugal. Despite the occurrence of scientific publications, we sought to understand the extent to which the Brazilian scientific community has addressed the theme of programming education at different  educational levels. For this purpose, we constructed a heat map (Figure~\ref{fig:map}) that correlated heat intensity with scientific publication density. ``High'' indicates high publication density, while ``Low'' denotes low density. Thus, to address $\mathcal{RQ}_{1}$, we verify a higher incidence of research in the southeastern and northeastern regions of Brazil.  

\begin{table}[!htbp]
\centering
\caption{Distribution of Publications by Region.}
\label{tab:publication_distribution}
\footnotesize
\begin{tabular}{lc}
\hline
\textbf{Region}              & \textbf{Overall Percentage}     \\ \hline
North                        & 10.34\%                         \\
Northeast                    & 27.59\%                         \\
Central-West                 & 13.79\%                         \\
Southeast                    & 28.74\%                         \\
South                        & 17.24\%                         \\
Portugal (Coimbra and Porto) & 2.30\%                          \\ \hline
\end{tabular}
\end{table}

\begin{figure}[!htbp]
    \centering
    \includegraphics[width=0.6\linewidth]{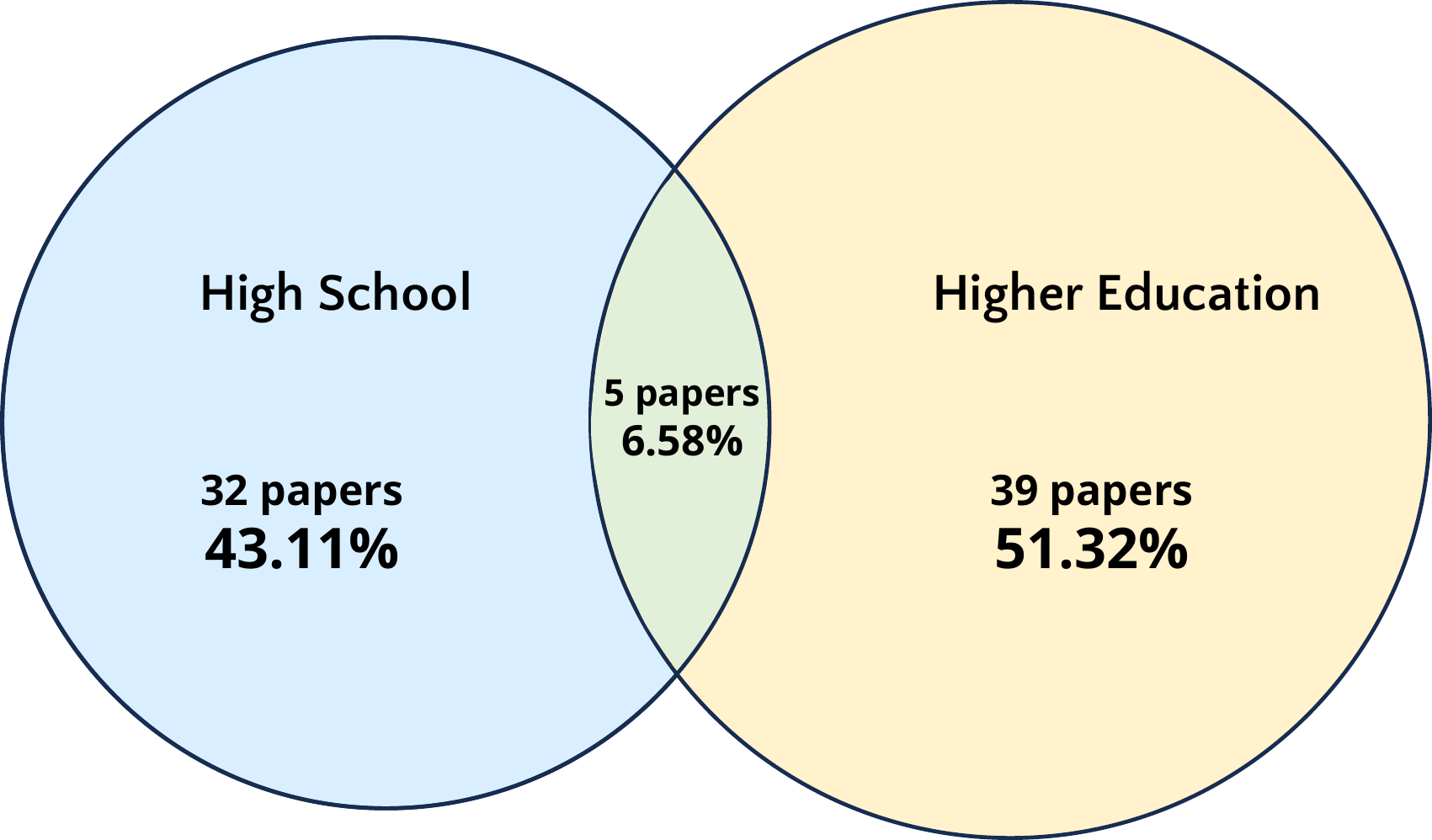}
    \caption{Number and percentage of publications by educational level.}
    \label{fig:venn}
\end{figure}

We found that approximately 43.11\% of the published works were conducted in high school, and 51.32\% in higher education. Interestingly, only 6.58\% of the studies addressed both educational levels, as shown in Figure~\ref{fig:venn}. This distribution suggests  heightened attention to higher education, potentially in response to dropout challenges encountered at this academic level.

\begin{figure}[!htbp]
    \centering
    \includegraphics[width=0.8\linewidth]{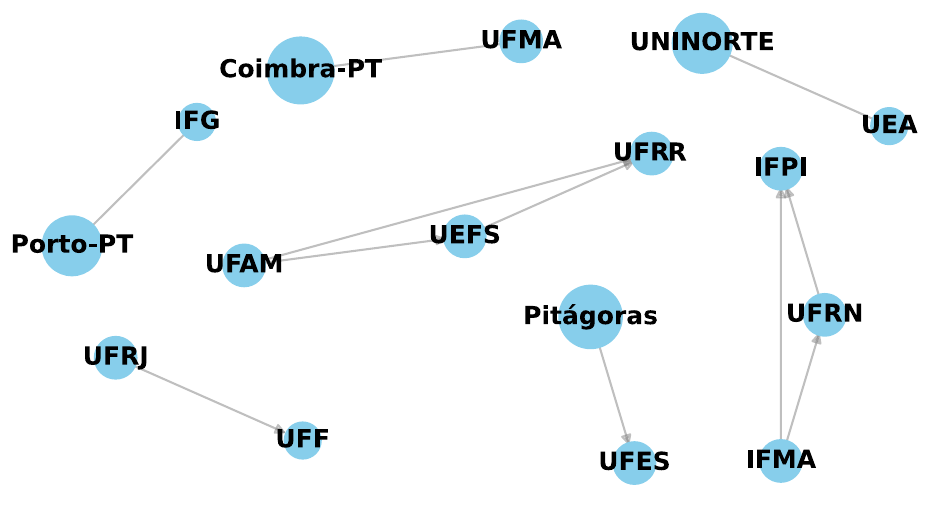}
    \caption{Institutions with publications in collaboration.}
    \label{fig:publications_partnership}
\end{figure}

Finally, we analyzed the  collaborations between institutions in research efforts, as illustrated in Figure~\ref{fig:publications_partnership}. Each node represents an institution and the edges between them denote the collaborative relationships established in the context of programming teaching studies. In addition, our analysis revealed that the majority of collaborations occur between institutions in the Northeast and North regions of Brazil, as exemplified by works published by~\cite{Silva2016, Braz2021}. Furthermore, we observed two international collaborations involving institutions from Portugal and institutions from the Northeast and Midwest regions of Brazil~\cite{Martins2019, Cruz2022}.

\textbf{$\mathcal{RQ}_{2}$: Which methods have been employed in programming teaching proposals?}


In response to $\mathcal{RQ}_{2}$ regarding the methods used in programming teaching across different education levels in Brazil, our analysis revealed that methodologies based on games and robotics are predominantly employed. This understanding is derived from Table~\ref{tab:methods}, which shows that 18.67\% (game) and 16\% (robotic) of the surveyed studies report this method to be used (\cite{Reis2019, Stephan2020, Silva2022}). In the other methods, we included 24 approaches, such as active methodologies, gamification, and exams, each representing 1.33\% of the studies.   


\begin{table}[!htbp]
\caption{Methodologies usage.}
\label{tab:methods}
\centering
\footnotesize
\begin{tabular}{lc}
\hline
\multicolumn{1}{c}{\textbf{Method}} & \textbf{Overall Percentage} \\ \hline
Game                                & 18.67\%                     \\
Robotic                             & 16\%                        \\
Extracurricular Classes             & 8\%                         \\
Problem Based Learning (PBL)        & 8\%                         \\
Unplugged Computation               & 4\%                         \\
Collaborative Learning               & 2.67\%                      \\
Workshop                            & 2.67\%                      \\
Others                              & 1.33\%                      \\ \hline
\end{tabular}
\end{table}


\textbf{$\mathcal{RQ}_{3}$: Which programming content should be addressed the most?}

\begin{figure}[!htbp]
  \centering
  \includegraphics[width=0.59\linewidth]{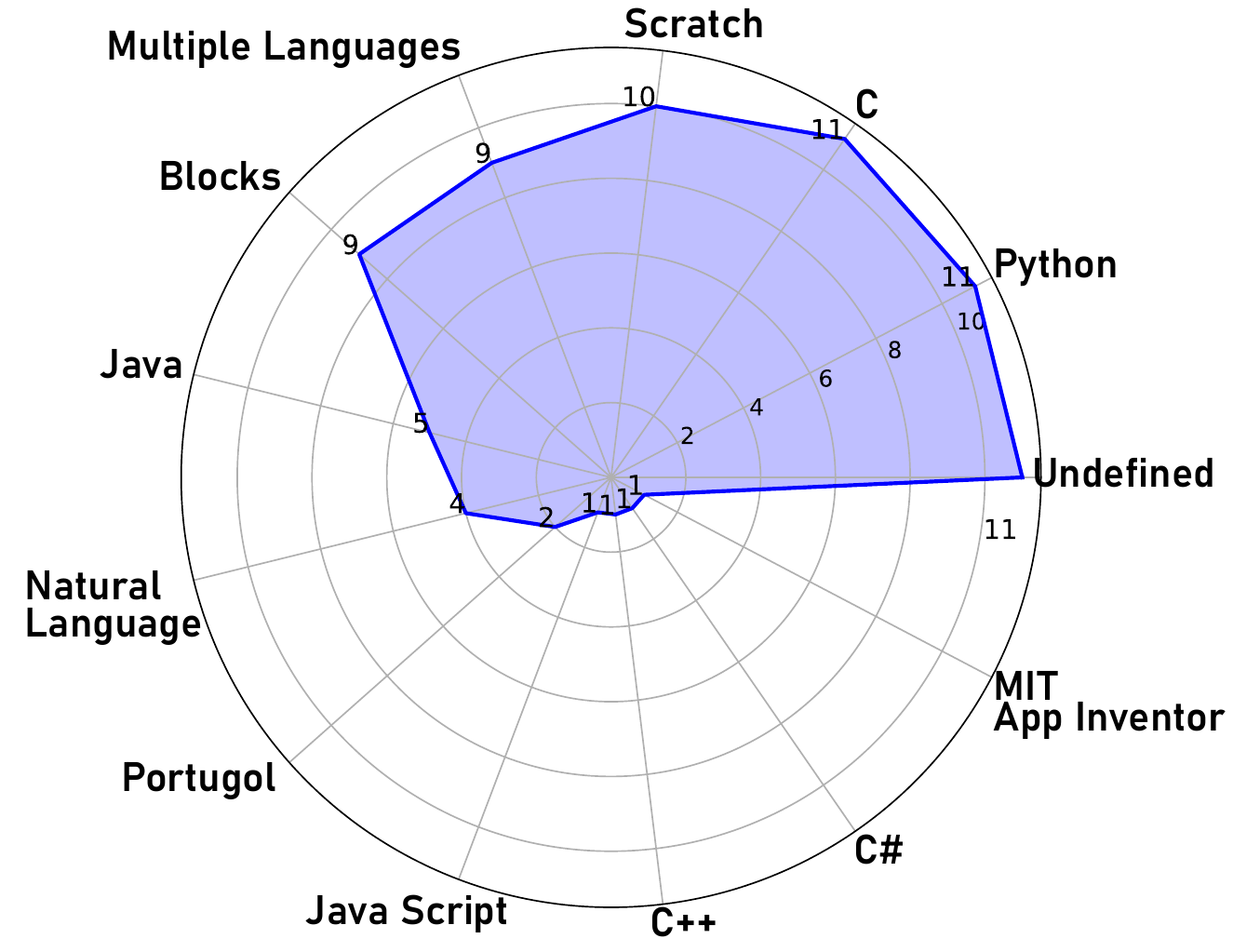}
  \caption{Programming language overview.}
  \label{fig:radar}
\end{figure}

To answer $\mathcal{RQ}_{3}$, we found that the most addressed content across different educational levels includes introductory programming concepts, such as variable declaration, decision structures, and looping structures, as evidenced by studies conducted by \cite{Jesus2019, Cruz2022, Gomes2023}. In addition, we found that the most commonly used programming languages were C and Python, as shown in Figure~\ref{fig:radar}.

Among the surveyed papers, it was observed that complex programming topics were not extensively covered in the studied methodologies, as exemplified by the teaching of pointers proposed by \cite{Kira2018} and the teaching of Data Structures in \cite{Roberto2018, Jesus2021}. However, it is still difficult to understand basic elements, such as programming logic, and this finding sheds light on which content could receive more attention in the development and improvement of teaching methodologies within the community.


\bigskip
\textbf{$\mathcal{RQ}_{4}$: How are programming teaching methods evaluated?}


In addition to the most frequently addressed content, it is crucial to understand the effectiveness of teaching methodology applications. In this regard, we surveyed the papers to address $\mathcal{RQ}_{4}$. We found that among the various ways to assess the effectiveness of the teaching method, the authors predominantly employed quantitative and qualitative forms. 

Through these forms, commonly reported findings relate to the quantitative analysis of student performance before and after the implementation of the teaching methodology. To a lesser extent, some authors have used more sophisticated approaches to evaluate the effectiveness of the applied methodology in programming teaching~\cite{Junior2021} noticed that employing their method, they achieved more uniform development of the class, reducing the heterogeneity and improving the engagement and interest of the students in the course content.

\bigskip
\textbf{$\mathcal{RQ}_{5}$: How does the programming language integrate into the programming teaching method?}

\begin{figure}[!htbp]
    \centering
    \centering
    \includegraphics[width=0.3\linewidth]{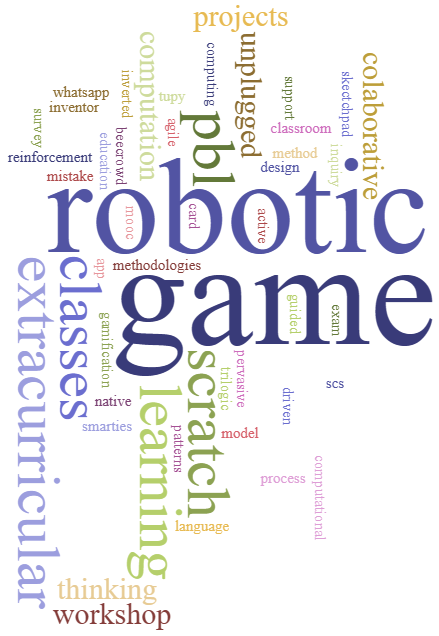}
    \caption{Teaching methodology word count.}
    \label{fig:word_cloud}
\end{figure}

We discretized and evaluated the relationship between programming language and programming teaching method using Pearson correlation and concluded that there is a weak correlation between programming language and teaching method, approximately 0.14. Additionally, we assessed the teaching methods used in programming education, as shown in Figure~\ref{fig:word_cloud}. 

\begin{figure}[!htbp]
    \centering
    \includegraphics[width=0.45\linewidth]{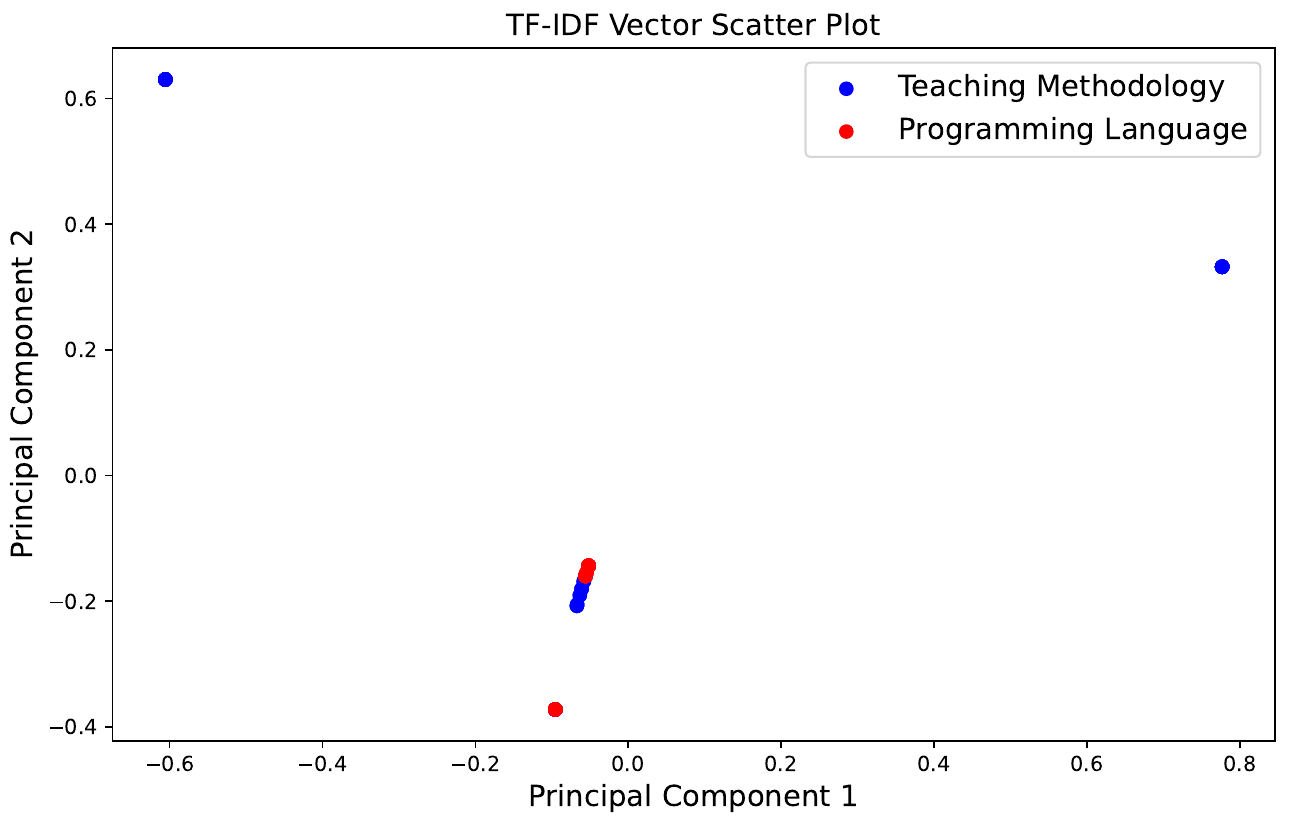}
    \caption{Relationship between method texts using TF-IDF and PCA.}
    \label{fig:word_correlation}
\end{figure}

We explored this relationship using Term Frequency-Inverse Document Frequency (TF-IDF), a text analysis technique that highlights the most significant terms by assigning weights to common words in a document, but rarely in the document set. Through Principal Component Analysis (PCA), we reduced the dimensionality of the TF-IDF vectors, capturing the main directions of variation in the data.
In Figure~\ref{fig:word_correlation}, each dot represents text with a color indicating the category. The axes depict the main directions of variation in the data, enabling the identification of patterns and trends in texts. PCA 1 and 2 represent the directions of the maximum variation after reduction. PCA 1 reflects the most differential words among the texts, whereas PCA 2 captures additional information not explained by PCA 1, providing a comprehensive view of word relationships. This highlights the effectiveness of Computational Thinking techniques. Shedding light on these results in response to $\mathcal{RQ}_{5}$, we find that despite the consistent work on the Computational Thinking theme, research challenges persist.

In addition, we summarize in Table~\ref{tab:summary_RQs} all raised $\mathcal{RQ}s$ and the findings.

\begin{table}[!htbp]
\centering
\renewcommand{\arraystretch}{1.1}
\caption{Research questions and summary findings.}
\label{tab:summary_RQs}
\resizebox{\textwidth}{!}{%
\begin{tabular}{ll}
\hline
\multicolumn{1}{c}{\textbf{Research Question ($\mathcal{RQ}$)}}                                                                                       & \multicolumn{1}{c}{\textbf{Finding}}                                                                                                                                                                                          \\ \hline
\begin{tabular}[c]{@{}l@{}} $\mathcal{RQ}_{1}$ \\ What are the overview of studies \\in Brazil?\end{tabular}                                              & \begin{tabular}[c]{@{}l@{}}Greater number of publications at  WEI and WIE conferences. \\ More papers published by institutions from the Southeast and Northeast.\\ More research in Higher Education (51.32\%).\end{tabular} \\ \hline
\begin{tabular}[c]{@{}l@{}}$\mathcal{RQ}_{2}$ \\ Which methods have been employed \\in programming teaching proposals? \end{tabular}                & Robotic, Game, and Extracurricular Classes.                                                                                                                                                                                   \\ \hline
\begin{tabular}[c]{@{}l@{}} $\mathcal{RQ}_{3}$ \\ Which programming content \\should be addressed the most?\end{tabular}                                  & Decision structures ($if$-$else$, $switch$) and looping structures ($for, while$).                                                                                                                                                            \\ \hline
\begin{tabular}[c]{@{}l@{}}$\mathcal{RQ}_{4}$ \\ How are programming teaching \\ methods evaluated?\end{tabular}                                   & Quantitative and qualitative forms.                                                                                                                                                                                           \\ \hline
\begin{tabular}[c]{@{}l@{}}$\mathcal{RQ}_{5}$ \\ How does the programming language \\integrate into the programming \\teaching method?\end{tabular} & \begin{tabular}[c]{@{}l@{}} It is decoupled, being the result of successful research in Computational Thinking.\end{tabular}                                                                                                \\ \hline
\end{tabular}}
\end{table}

\section{Conclusion}\label{sec:concluding_remark}

This study presents the research landscape in Brazil on Programming Education in high school and higher education from 2013 to 2023. We conducted a systematic review analyzing studies from the SBC-Open-Lib source to comprehend and address research inquiries pertinent to programming teaching. Our literature analysis sheds light on specific insights, including the primary coding languages employed, methods utilized, and target domains. Understanding these aspects empowers decision makers to construct robust and contemporary national education policies. 

In addition to our findings, we observed that robotics and gamming are widely used to compute programming teaching methods. In future work, we plan to evaluate the long-term effects of these methodologies on student learning and career preparedness. Overall, our future endeavors will seek to advance programming education in Brazil by fostering an environment that empowers all students to excel in the field of technology.

\section*{Acknowledgments}
We would like to thank PIBEN/FUNARBEN, CAPES and FAPESP MCTIC/CGI Research project 2018/23097-3 for the financial support. 
Andr\'e R. Backes gratefully acknowledges the financial support of CNPq (Grant \#307100/2021-9). 
This study was financed in part by the Coordenação de Aperfeiçoamento de Pessoal de Nível Superior - Brasil (CAPES) - Finance Code 001.

%
%
\bibliography{references}

\end{document}

%% file: acronym.tex
\acrodef{3GPP}{3rd Generation Partnership Project}
\acrodef{AI}{Artificial Intelligence}
\acrodef{B5G}{Beyond Fifth Generation}

\acrodef{CUBIC}{Conjunctive Using BIC (Binary Increase Congestion Control)}
\acrodef{cwnd}{Congestion Window}
\acrodef{DoS}{Denial of Service}
\acrodef{DDoS}{Distributed Denial of Service}
\acrodef{DNN}{Deep Neural Network}
\acrodef{DRL}{Deep Reinforcement Learning}
\acrodef{DT}{Decision Tree}
\acrodef{DNN}{Deep Neural Network}
\acrodef{DMP} {Deep Multilayer Perceptron}
\acrodef{DQN}{Deep Q-Learning}
\acrodef{ETSI}{European Telecommunications Standards Institute}
\acrodef{FIBRE}{Future Internet Brazilian Environment for Experimentation}
\acrodef{FTP}{File Transfer Protocol}
\acrodef{Flat}{Flat Neural Network}
\acrodef{GNN}{Graph Neural Networks}
\acrodef{HTM}{Hierarchical Temporal Memory}

\acrodef{IAM}{Identity And Access Management}
\acrodef{IID}{Informally, Identically Distributed}
\acrodef{IoE}{Internet of Everything}
\acrodef{IoT}{Internet of Things}
\acrodef{KNN}{K-Nearest Neighbors}
\acrodef{LSTM}{Long Short-Term Memory}
\acrodef{MPTCP}{Multipath Transmission Control Protocol}
\acrodef{M2M}{Machine to Machine}
\acrodef{MAE}{Mean Absolute Error}
\acrodef{ML}{Machine Learning}
\acrodef{MOS}{Mean Opinion Score}
\acrodef{MAPE}{Mean Absolute Percentage Error}
\acrodef{MSE}{Mean Squared Error}
\acrodef{mMTC}{Massive Machine Type Communications}
\acrodef{MFA}{Multi-factor Authentication}
\acrodef{MQTT}{Message Queuing Telemetry Transport}

\acrodef{NN}{Deep Neural Network}
\acrodef{NS3}{Network Simulator 3}
\acrodef{OSM}{Open Source MANO}
\acrodef{QL}{Q-learning}
\acrodef{QoE}{Quality of experience}
\acrodef{QoS}{Quality of Service}
\acrodef{RAM}{Random-Access Memory}
\acrodef{RF}{Random Forest}
\acrodef{RL}{Reinforcement Learning}
\acrodef{RMSE}{Root Mean Square Error}
\acrodef{RNN}{Recurrent Neural Network}
\acrodef{Reno}{Regular NewReno}
\acrodef{RTT}{Round Trip Time}
\acrodef{SDN}{Software-Defined Networking}
\acrodef{SFI2}{Slicing Future Internet Infrastructures}
\acrodef{SLA}{Service-Level Agreement}
\acrodef{SON}{Self-Organizing Network}

\acrodef{TCP}{Transmission Control Protocol}
\acrodef{VoD}{Video on Demand}
\acrodef{VR}{Virtual Reality}
\acrodef{V2X}{Vehicle-to-Everything}
